\documentstyle[prb,aps,twocolumn,epsfig]{revtex}
\newcommand{\be}{\begin{eqnarray}}
\newcommand{\ee}{\end{eqnarray}}

\begin{document}
\draft
\twocolumn[\hsize\textwidth\columnwidth\hsize\csname @twocolumnfalse\endcsname

\title{Optical Studies of a Layered Manganite La$_{1.2}$Sr$_{1.8}$Mn$_2$O$_7$ :
Polaron Correlation Effect}
\author{H. J. Lee, K. H. Kim, J. H. Jung, and T. W. Noh}
\address{Department of Physics and Center for Strongly Correlated Materials Research,\\
Seoul National University, Seoul 151-742, Korea}
\author{R. Suryanarayanan, G. Dhalenne, and A. Revcolevschi}
\address{Laboratoire de Physico Chimie de l'Etat Solide CNRS, UMR8648, B\^{a}t, 414,\\
Universit\'{e} Paris-Sud, 91405 Orsay, France}
\date{\today }
\maketitle

\begin{abstract}
Optical conductivity spectra of a cleaved {\it ab}-plane of a La$_{1.2}$Sr$_{1.8}$Mn$_2$O$_7$ single crystal exhibit a small polaron absorption band in
the mid-infrared region at overall temperatures. \ With decreasing
temperature ($T$) to Curie temperature ($T_C$), the center frequency of the
small polaron band moves to a higher frequency, resulting in a gap-like
feature, and that it collapses to a lower frequency below $T_C$. \
Interestingly, with decreasing $T$, the stretching phonon mode hardens above 
$T_C$ and softens below $T_C$. These concurring changes of lattice and
electronic structure indicate that short range polaron correlation exist
above $T_C$ but disappear with a magnetic ordering.
\end{abstract}


\vskip1pc] \newpage

Recent studies on manganites have shown that there exist strong coupling
among spin, charge, orbital, and lattice degrees of freedom. The relative
coupling strength of those degrees of freedom can be sensitively affected by
variation of physical parameters, such as amounts of carrier doping, and/or
structural modification. For example, the structure of cubic perovskite
(La,Sr)MnO$_3$ can be modified into a layered one by inserting a
rock-salt-type block layer (La,Sr)$_2$O$_2$ into every {\it n}-MnO$_2$
sheets, i.e., by forming the Ruddlesden-Popper compound, (La,Sr)$_{n+1}$Mn$%
_n $O$_{3n+1}$. With the variation of structures from single$-$ ({\it n}=1: K%
$_2 $NiF$_4$ structure), double$-$ ({\it n}=2) and $\infty -$ (cubic
perovskite) MnO$_2$ sheet, physical properties of these materials are
sensitively varying.\cite{moritomo} In addition, the effective low
dimensionality of the reduced {\it n} system can enhance charge and spin
fluctuations to induce more localized tendency than the cubic one. \ 

La$_{1.2}$Sr$_{1.8}$Mn$_2$O$_7$, which has the double MnO$_2$ sheets, is a
prototypical material that exibits intriguing interplays of various degrees
of freedom. Although it becomes a ferromagnetic (FM) metal below Curie
temperature ($T_C$) at $126$ K, earlier studies showed significant local
Jahn-Teller (J-T) lattice distortions at overall temperatures\cite{mitchell}
and short range antiferromagnetic spin order.\cite{perring} A more recent
study also provided a clear evidence of lattice polaron formation above $T_C$
by showing diffuse X-ray scattering around the Bragg peaks. At the same
time, the scattering experiments indicated an existence of short range
polaron ordering by showing incommesurate satellite peaks.\cite{doloc} \ 

Optical spectra for this bilayered system have been reported already.\cite
{ishikawa} However, there are no systematic optical investigations how
polaron effects with short range correlation become manifest in the optical
spectra. In this report, we present detailed optical conductivity spectra
which reveal interplays of spin, charge, and lattice degrees of freedom in La%
$_{1.2}$Sr$_{1.8}$Mn$_2$O$_7$. With decreasing $T$, the mid-infrared small
polaron band moves to a higher frequency up to $T_C$ and then becomes
collapsed to a lower frequency below $T_C$. And, the stretching phonon mode
hardens above $T_C$ and softens below $T_C$. These concurring changes of
lattice and electronic structure support the existence of the enhanced
polaron (charge) correlation above $T_C$ and its sudden collapse below $T_C$%
. \ A single crystal of La$_{1.2}$Sr$_{1.8}$Mn$_2$O$_7$ was grown by the
floating-zone method using a mirror furnace. The sample was characterized by
resistivity and magnetization measurements.\cite{prellier} For optical
reflectivity measurements, a

\begin{figure}[tbp]
\epsfig{file=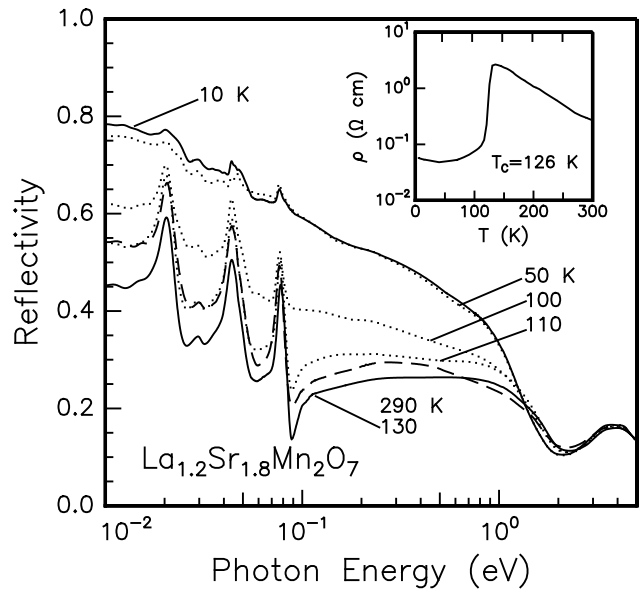, width=3.3in,clip=}
\vspace{2mm}
\caption{$T$-dependent $R(\omega )$ of $E\parallel ab$ for the La$_{1.2}$Sr$%
_{1.8}$Mn$_2$O$_7$ single crystal. Inset: in-plane $\rho (T)$ with $T_C$=126
K.}
\label{Fig:1}
\end{figure}

\begin{figure}[tbp]
\epsfig{file=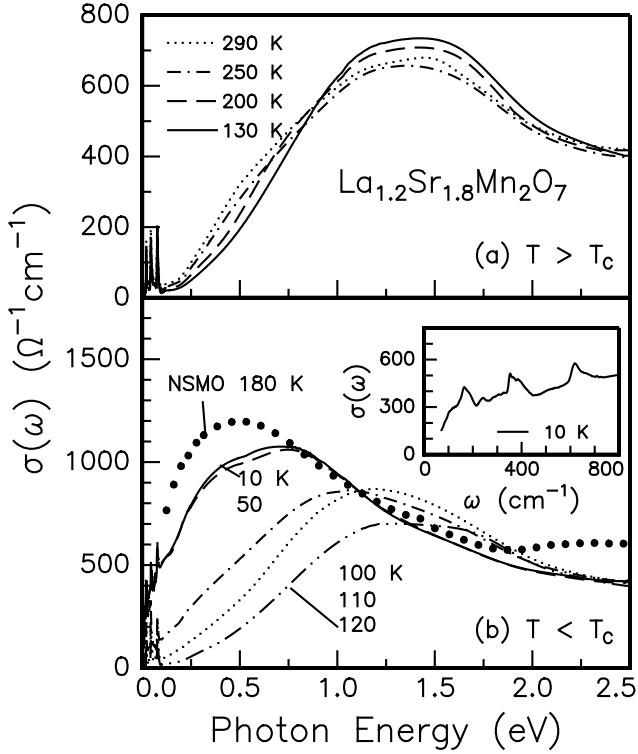, width=3.3in,clip=}
\vspace{2mm}
\caption{(a) $\sigma (\omega )$ of La$_{1.2}$Sr$_{1.8}$Mn$_2$O$_7$ for $%
T>T_C $ and (b) $T<T_C$. The solid circles in (a) and (b) represent the $%
\sigma (\omega )$ of Nd$_{0.7}$Sr$_{0.3}$MnO$_3$ at 180 K. Inset of (b) : $%
\sigma (\omega )$ of La$_{1.2}$Sr$_{1.8}$Mn$_2$O$_7$ in the low frequency
region at 10 K.}
\label{Fig:2}
\end{figure}

cleaved {\it ab}-plane was freshly prepared. Details for the reflectivity
measurements were described in our previous report.\cite{haej} Using the
Kramers-Kronig relation, we obtained optical conductivity spectra $\sigma
(\omega )$ from reflectivity spectra $R(\omega ) $.

Figure 1 shows $T$-dependent $R(\omega )$ of the cleaved {\it ab}-plane of La%
$_{1.2}$Sr$_{1.8}$Mn$_2$O$_7$. At 290 K, there are three sharp peaks
originating from optic phonon modes in the far-infrared region. With $T$
approaching $T_C$, $R(\omega )$ below 0.4 eV decrease, which is consistent
with the dc resistivity behavior shown in the inset of Fig. 1. As $T$
decreases further below $T_C$, $R(\omega )$ start to increase significantly,
apporaching to a metallic response.

$T$-dependent $\sigma (\omega )$ both above and below $T_C$ are displayed in
Figs. 2 (a) and (b), respectively. Above $T_C$, $\sigma (\omega )$ show a
broad absorption band around 1.2 eV. The shape of $\sigma (\omega )$ at 290
K looks similar to that at 250 K. However, as $T$ approaches 130 K, $\sigma
(\omega )$ below 0.5 eV become suppressed to show a finite gap-like feature.
At the same time, the height of the broad band near 1.2 eV increases to form
a sharper band. On the other hand, as $T$ is lowered below $T_C$, the
spectral weight moves suddenly to a lower energy region as shown in Fig. 2
(b). \ The shape becomes rather asymmetric and its magnitude below 1.0 eV
increases significantly, indicating that a large spectral weight become
transferred from a higher energy region with the onset of the magnetic
ordering.

\begin{figure}[tbp]
\epsfig{file=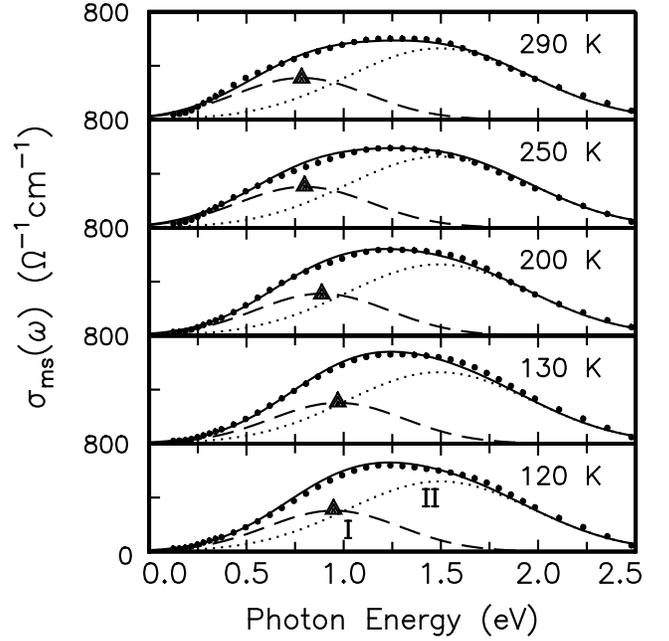, width=3.3in,clip=}
\vspace{2mm}
\caption{$\sigma _{ms}(\omega )$ of La$_{1.2}$Sr$_{1.8}$Mn$_2$O$_7$. The
solid circles are experimental data. The dotted lines represent fixed
Gaussian functions around 1.5 eV (Peak II). The dashed lines show small
polaron fitting using the Gaussian function (Peak I). The solid lines
represent the sums of two Gaussian functions. The solid triangles represent
the center of Peak I.}
\label{Fig:3}
\end{figure}

To get further insights, we compared the $T$-dependent $\sigma (\omega )$ of
La$_{1.2}$Sr$_{1.8}$Mn$_2$O$_7$ with those of Nd$_{0.7}$Sr$_{0.3}$MnO$_3$
(NSMO), \cite{haej} which is a cubic perovskite with a similar
metal-insulator transition. Compared with the other cubic perovskite
manganites such as La$_{0.7}$Ca$_{0.3}$MnO$_3$\cite{kim98} and La$_{0.7}$Sr$%
_{0.3}$MnO$_3$,\cite{okimoto} NSMO has a relatively low $T_C$ around $200$
K, due to a reduced electron bandwidth. It is now well established that
small polaron plays an important role in the paramagnetic insulating regime
of the perovskite manganites.\cite{noh} Furthermore, a recent X-ray and
neutron scattering studies on La$_{1.2}$Sr$_{1.8}$Mn$_2$O$_7$ showed the
existence of polarons in the paramagnetic state.\cite{doloc}\ Therefore, it
is quite reasonable that optical spectra of La$_{1.2}$Sr$_{1.8}$Mn$_2$O$_7$
above $T_C$ can be analyzed by the polaron picture.\cite{noh}

There exist some differences in the spectra of La$_{1.2}$Sr$_{1.8}$Mn$_2$O$%
_7 $ and NSMO. First, $\sigma (\omega )$ of NSMO and other cubic perovskites
were reported to be nearly $T$-independent above $T_C$.\cite
{haej,kim98,okimoto,noh,quijada} However, $\sigma (\omega )$ of La$_{1.2}$Sr$%
_{1.8}$Mn$_2$O$_7$ below 2 eV show a systematic $T$-dependence above $T_C$.
As $T$ approaches $T_C$, spectral weight below 1.0 eV decreases, but that
above 1.0 eV increases. Second, $\sigma (\omega )$ of La$_{1.2}$Sr$_{1.8}$Mn$%
_2$O$_7$ do not show a Drude-like peak even at 10 K (see the inset of Fig. 2
(b)). In the FM metallic states of the cubic perovskite manganites,\cite
{haej,kim98,okimoto,noh,quijada} the small polaron spectral weight was
transferred to a lower energy to form an asymmetric mid-infrared band and a
finite Drude-like peak at very low {\it T}, which were interpreted as
incoherent and coherent absorption bands of a large polaron, respectively.
The lack of the Drude peak in the $\sigma (\omega )$ of La$_{1.2}$Sr$_{1.8}$%
Mn$_2$O$_7$ can be related to its effective low dimensionality, induced by a
decrease of the number of the MnO$_2$ sheet (i.e. $n=2$). Third, the
lineshape of the La$_{1.2}$Sr$_{1.8}$Mn$_2$O$_7$ absorption band at 10 K $%
(\approx 0.1T_C$, $T_C=126$ K$)$ is quite similar to that of NSMO at 180 K $%
(\approx 0.9T_C$, $T_C=198$ K$)$, as shown in Fig. 2 (b). Note that $\sigma
(\omega )$ of NSMO at 180 K is close to that of 200 K (above $T_C$) in its
shape, without showing the Drude-like peak.\cite{haej} These observations
indicate that La$_{1.2}$Sr$_{1.8}$Mn$_2$O$_7$ remain as the small polaron
state even far below $T_C$.\ Therefore, it is reasonable that the $\sigma
(\omega )$ of La$_{1.2}$Sr$_{1.8}$Mn$_2$O$_7$ can be analyzed in the small
polaron picture at all temperatures.

To get a more quantitative information on the small polaron absorption, we
analyzed the experimental $\sigma (\omega )$ in terms of $\sigma (\omega
)=\sigma _{ms}(\omega )+\sigma _L(\omega )$.\cite{jung98} Here, $\sigma
_{ms}(\omega )$ represent the conductivity contribution of the two mid-gap
states below 2.0 eV and $\sigma _L(\omega )$ correspond to the
charge-transfer transition between $O$ 2{\it p }and Mn $e_g$ levels,
centered around 4.0 eV. By fitting with a Lorentz oscillator, we determined $%
\sigma _L(\omega )$ and subtracted it from the experimental $\sigma (\omega
) $ to obtain $\sigma _{ms}(\omega )$. For fitting $\sigma _{ms}(\omega )$,
we used two Gaussian functions as shown in Fig. 3.\cite{alex} One is located
below 1.0 eV (Peak I) and the other is around 1.5 eV (Peak II).\cite
{machida,allen} Peak I corresponds to the small polaron absorption related
to a nearest neighbor hopping from Mn$^{3+}$ to Mn$^{4+}$,\cite
{haej,kim98,okimoto,noh,quijada} and Peak II corresponds to on-site $d-d$
transition.\cite{dd} Interestingly, $\sigma _{ms}(\omega )$ above and just
below $T_C$ can be well described when only the center frequency of Peak I, $%
\omega _{\text{I}}$, is assumed to be $T$ -dependent, while the other
parameters such as the strength and the width of Peak I are fixed. And, Peak
II are nearly $T$ -independent within 3 \%. Fig. 3 shows the fitting results
above and just below $T_C$. With lowering $T$ further, the best fitting was
obtained when the strength of Peak I as well as\ $\omega _{\text{I}}$ was
assumed to change with a slight variation of the strength of Peak II. \ 

Fig. 4 (a) shows $T$-dependence of $\omega _{\text{I}}$ obtained by the
fitting process. As $T$ becomes lower in the paramagnetic region, $\omega _{%
\text{I}}$ clearly increases from 0.8 to 1.0 eV. With magnetic ordering at $%
T_C$, $\omega _{\text{I}}$ starts to decrease abruptly to reach a finite
value of 0.58 eV at 10 K. In case of an adiabatic small polaron, $\omega _{%
\text{I}}$ corresponds to two times of the small polaron binding energy.\cite
{emin} Therefore, above results indicate that the coupling between charge
and lattice should exist far above $T_C$ and that its strength be enhanced
near $T_C$. And, the coupling strength suddenly decreases to a lower value
with the influence of spin ordering. \ \ 

The increase of $\omega _{\text{I}}$ at the high $T$ region is responsible
for the apparent suppression of $\sigma (\omega )$ below 0.4 eV, shown in
Fig. 2 (a). The suppression of the $\sigma (\omega )$ produces a finite
gap-like tail below 0.4 eV. The tail moves systematically to a higher energy
between 250 K and $T_C$ and the gap-like behavior becomes evident around 130
K just above $T_C$. This behavior is reminiscent of a finite

\begin{figure}[tbp]
\epsfig{file=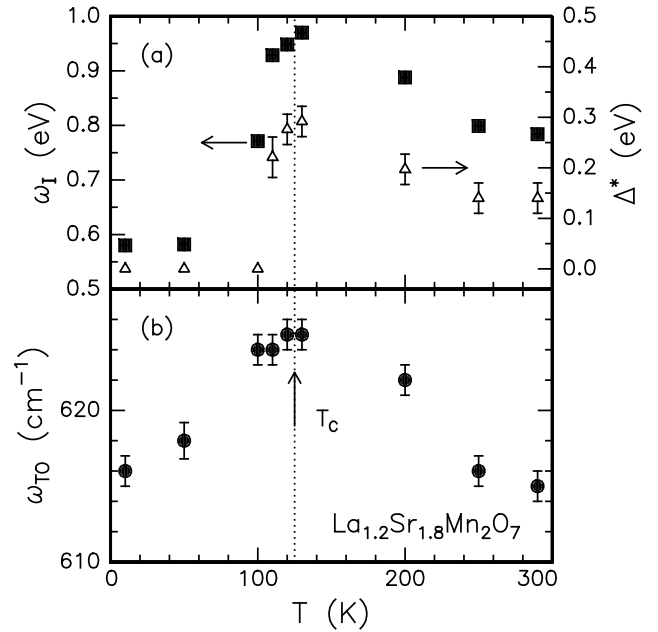, width=3.3in,clip=}
\vspace{2mm}
\caption{(a) $T$-dependent center frequencies of Peak I ($\omega _{\text{I}}$%
) and $\Delta ^{*}$ [see the main text for definition] are represented by
solid square and open triangle, respectively. Error bars in $\omega _{\text{I%
}}$ are smaller than the size of solid square. (b) The peak frequencies of
the stretching phonon mode ($\omega _{\text{TO}}$). The dotted line is a
guide to depict $T_C$. }
\label{Fig:4}
\end{figure}

charge gap formation in the materials with a clear long range charge
ordering (CO) at low $T$.\cite{jung00,liu} Because there is no evidence for
the long range CO in La$_{1.2}$Sr$_{1.8}$Mn$_2$O$_7$, the peculiar behavior
may suggest an existence of short range charge correlation above $T_C$. In
reality, a recent scattering experiment confirmed the existence of this
short range charge and polaron correlation that grows up near $T_C$ and
diminishs below\ $T_C$.\cite{doloc} To quantify the short range charge order
above $T_C$, we define $\Delta ^{*}$ as a crossing point energy with the $%
\sigma (\omega )=0$ line when a steeply increasing part of $\sigma (\omega )$
is linearly extrapolated. Fig. 4 (a) summarizes the $T$-dependent values of $%
\Delta ^{*}$. The $T$-dependence of $\Delta ^{*}$ is quite similar to that
of $\omega _{\text{I}}$. With decreasing $T$ to $T_C$, $\Delta ^{*}$
gradually increases from 0.15 to 0.28 eV. The $\Delta ^{*}$ suddenly starts
to decrease at $T_C$ and\ becomes zero below $\sim $100 K. These
experimental results of $\omega _{\text{I}}$ and $\Delta ^{*}$ strongly
support that polaron and charge correlations grow up to $T_C$ and collapse
due to the FM ordering. \ 

The stretching optical phonon mode, related to the lattice degree of
freedom, also reflects the existence of short range charge and polaron
correlations above $T_C$. Figure 5 presents $T$-dependence of the phonon
mode $\omega _{\text{TO}}$ located around 615 cm$^{-1}$ at 290 K. With
lowering $T$ to $T_C$, $\omega _{\text{TO}}$ shows a significant hardening.\
When $T$ is further lowered, the phonon mode softens.\ Figure 4 (b) shows
the values of $\omega _{\text{TO}}$ determined by fitting with the Lorentz
oscillator. Between 290 and 130 K, $\omega _{\text{TO}}$ increases by about
10 cm$^{-1}$. With lowering $T$, it is clearly shown that $\omega _{\text{TO}%
}$ decreases abruptly to 616 cm$^{-1}$ at 10 K. 
\begin{figure}[tbp]
\epsfig{file=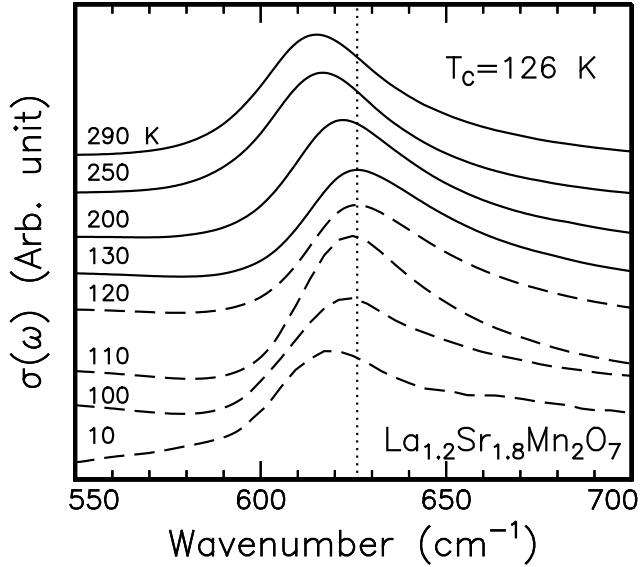, width=3.3in,clip=}
\vspace{2mm}
\caption{$T$-dependence of the optical stretching phonon modes. The solid
and the dashed lines are phonon modes above and below $T_C$, respectively.
The dotted line represents the peak frequency of the phonon mode around 625
cm$^{-1}$ at 130 K. }
\label{Fig:5}
\end{figure}

The frequency shift of the stretching mode reflects that there exist
significant modulations of local Mn-O bond lengths.\cite{kim96}\ In long
range CO systems, the stretching phonon mode hardens near CO temperature, $%
T_{\text{CO}}$: the observed frequency shifts were about 15 cm$^{-1}$ and 25
cm$^{-1}$ in case of La$_{0.5}$Ca$_{0.5}$MnO$_3$\cite{kim-phd} and Bi$%
_{0.18} $Ca$_{0.82}$MnO$_3$,\cite{liu} respectively. The hardening behavior
of the stretching mode in La$_{1.2}$Sr$_{1.8}$Mn$_2$O$_7$ above $T_C$ is
somewhat similar to that observed in La$_{0.5}$Ca$_{0.5}$MnO$_3$ and Bi$%
_{0.18}$Ca$_{0.82}$MnO$_3$ near $T_{\text{CO}}$. This observation is also
consistent with the results of Fig. 4 (a), showing the existence of charge
correlation effects in La$_{1.2}$Sr$_{1.8}$Mn$_2$O$_7$.\ In addition, the
abrupt softening of $\omega _{\text{TO}}$ below $T_C$ should be understood
in terms of the melting of the short range spatial correlation influenced by
the spin ordering.

All our experimental findings suggest that there should be intimate coupling
among charge, spin, and lattice degrees of freedom (through polaron) and
that they interplays with each other in La$_{1.2}$Sr$_{1.8}$Mn$_2$O$_7$.
Especially, short range spatial correlation effects of the charge, spin, and
lattice degrees of freedom can be used to explain the $T$-dependence of $%
\sigma (\omega )$, such as the gap-like behaviors and increase of small
polaron peak frequency in the mid-infrared region.\ [And, the dynamic
fluctuations of those various degrees of freedom can be also important in a
similar {\it T} window.]

In summary, optical conductivities in La$_{1.2}$Sr$_{1.8}$Mn$_2$O$_7$
indicate that the short range correlation of polarons exist above $T_C$, and
that the sample remains as small polaron state even at 10 K. Subtle balance
and competition among the spin, charge, and lattice degrees of freedom
should be considered in understanding optical properties of the layered
manganite.

We thank to H. K. Lee, Y. S. Lee, and Dr. Y. Chung for useful discussion and
helpful experiments. This work was supported by Ministry of Science and
Technology through the Nanostructure Technology Project and by the BK-21
Project of the Ministry of Education.

\end{document}